\documentclass[twocolumn]{revtex4}
\pdfoutput=1
\usepackage{latexsym}
\usepackage{amsmath}
\usepackage{times}
\usepackage{amssymb}
\usepackage{fancyheadings}
\usepackage[T1]{fontenc}
\usepackage[utf8]{inputenc}
\usepackage{url}
\usepackage{hyperref}
\usepackage{color}
\usepackage{graphicx}
 \usepackage{epsfig}
\usepackage{mathptmx}
\usepackage{bm}
\usepackage{array}
\usepackage{algorithm}
\usepackage{algorithmic}
\usepackage{url}
\usepackage{hyperref}
\usepackage{algorithm}
\usepackage{algorithmic}
\usepackage{color}

\usepackage{lscape}

\usepackage{draftcopy}
\begin{document}
\title{Dynamical implications of sample shape for avalanches in 2-dimensional random-fiel Ising model with saw-tooth domain wall}
\author{Bosiljka Tadi\'c}
\affiliation{ %\flushleft
Department  for Theoretical Physics; Jo\v{z}ef Stefan Institute; 
P.O. Box 3000; SI-1001 Ljubljana; Slovenia\\ \hspace{1cm}} 

%\date{}
\begin{abstract} 
We study dynamics of a  built-in domain wall (DW) in 2-dimensional disordered ferromagnets with different sample shapes using random-field Ising model on a square lattice rotated by  45 degrees. The saw-tooth DW of the length $L_x$ is created along one side and swept through the sample by slow ramping of the external field until the complete magnetisation reversal and the wall annihilation at the open top boundary at a distance  $L_y$. By fixing the number of spins $N=L_x\times L_y=10^6$ and the random-field distribution at a value above the critical disorder, we vary the ratio of the DW length to the annihilation distance in the range $L_x/L_y\in[1/16,16]$. The periodic boundary conditions are applied in the y-direction so that these ratios comprise different samples, i.e., surfaces of cylinders with the changing perimeter $L_x$ and height $L_y$. We analyse the avalanches of the DW slips between following field updates, and the multifractal structure of the magnetisation fluctuation time series.
Our main findings are that the domain-wall lengths materialised in
different sample shapes have an impact on the dynamics at all scales. Moreover, the domain-wall motion at the beginning of the hysteresis loop (HLB) probes the disorder effects resulting in the fluctuations that are significantly different from the large avalanches in the central part of the loop (HLC), where the strong
fields dominate. Specifically, the fluctuations in HLB exhibit a wide multi-fractal spectrum, which shifts towards higher values of the exponents when the DW length is reduced. The distributions of the avalanches in this segments of the loops obey power-law decay and the exponential cutoffs with the exponents firmly in the mean-field universality class for long DW. In contrast, the avalanches in the HLC obey Tsallis density distribution with the power-law tails which indicate the new categories of the scale invariant behaviour for different ratios $L_x/L_y$.
 The large fluctuations in the HLC, on the other hand, have a rather narrow spectrum which is less sensitive to the length of the wall. These findings shed light to the dynamical criticality of
the random-field Ising model at its lower critical dimension; 
they can be relevant to applications of the dynamics of
injected domain walls in two-dimensional nanowires and ferromagnetic films.
\end{abstract}

\maketitle

 \section{Introduction}
Controlled injection and motion of domain walls in planar magnetic wires and ferroic nanowires are of a renewed interest for the
applications in new memory devices and domain-wall logic 
\cite{DW-logic_review,science-rew2008,NatComm2015_White_DwFE}. Techniques to inject a domain wall have been developed in technologically interesting materials as ferromagnetic films \cite{koreans_JApplPhys2008} and planar Permalloy nanowires
\cite{DWinject_NW2002,saveljev_NJP2005,DWinject_NL2015} as well as in ferroelectric thin films \cite{NatComm2015_White_DwFE,NatNanotech2015_Tagantsev_DwFE}. 
The motion of the wall in these systems results from an interplay between the applied external field and pinning at intrinsic defects and the morphology of the sample boundary. Some results indicate that the shape of the sample becomes increasingly relevant with the reduced dimensionality \cite{DW_NWshape2017}.  
Moreover, the new visualisation techniques provide a
direct observation of the domain-wall motion in different experimental settings \cite{AFM_BHN2004,koreans_NatPhys2007,koreans_JApplPhys2008,DWstochasticity_NatComm2010,DWstochasticity_expresistanceNoise2010,DWstochasticity2011,koreans2017}; these experiments disclose important details of the stochastic dynamics and avalanching behaviour. For instance,  the studies in 
\cite{koreans_NatPhys2007,koreans_JApplPhys2008} reveal the
domain-evolution patterns in ferromagnetic MnAs and Co films using time-resolved magneto-optical microscopy. The observed avalanches of domain-wall slips near the coercive field obey the
scale invariance with tuneable scaling exponents, which depend on the position of the present domain wall concerning the underlying lattice, varied by changing temperature. 
In this context, theoretical investigations can provide additional information about the nature of the underlying stochastic processes of magnetisation reversal in two-dimensional disordered ferromagnets and narrow stripes with a built-in domain wall.

Disordered ferromagnetic films exhibiting criticality at the
hysteresis loop \cite{koreans_JApplPhys2008,pupin2000,koreans_NatPhys2007,koreans_films2011,koreans_films2013} are conveniently modeled by  two-dimensional random-field Ising model (2D-RFIM) \cite{djole_PRL_critdisorder2011,djole_scaling2D,djole_spanningaval2014}, as described in Section\ \ref{sec:model}.
Besides, these model systems are also of high importance for theoretical considerations for the following reasons. First,  given that $d=2$ appears to be the lower critical dimension of RFIM \cite{aizeman1989}, there are no equilibrium phase
transitions in 2D-RFIM  (see a more detailed discussion in
\cite{djole_scaling2D}). However, the non-equilibrium 2D-RFIM has a rich critical behaviour at the hysteresis loop with scale-invariance of the Barkhausen avalanches, as recently demonstrated by the extensive simulations in \cite{djole_PRL_critdisorder2011,djole_scaling2D}. 
Furthermore, compared to often studied three-dimensional model
\cite{eduard_FSS2003,eduard_spanning2004,BT_EPJB2002},  the
avalanching dynamics in 2D-RFIM appear to be more sensitive to the effects of the strength of the random-field disorder
\cite{djole_PRL_critdisorder2011,djole_spanningaval2014}, the presence of nonmagnetic defects \cite{BT_PRL1996}, and for boundary conditions \cite{BT_PRE2000}. 
For instance, the finite-size scaling analysis using a vast span
of system sizes \cite{djole_PRL_critdisorder2011,djole_spanningaval2014} revealed that the probability of spanning avalanches vanishes at a rather small critical disorder $R_c=0.54\pm0.06$, compared to $R_c=2.16$ in the three-dimensional model. 
Recently, the multi-fractal analysis of Barkhausen noise in 3D-RFIM  demonstrated \cite{BT_MFRbhn2016}  a broad range of temporal scales in the magnetisation reversal processes. Although a non-trivial fractal structure can be expected for the Barkhausen noise in the 2D-RFIM, a detailed theoretical study is still missing.

In this work, we study the dynamics of a built-in  domain wall in
2D-RFIM on a square lattice rotated by $45^0$; in this configuration, as shown in the Inset to Figure\ \ref{fig-schema}, the saw-tooth wall
which is prepared along the sample boundary of the length $L_x$ can be moved by an infinitesimal change in the field when the disorder is absent. Therefore, by slowly ramping the external field along the descending branch of the hysteresis loop, the motion of the domain-wall at the beginning of the loop probes the true effects of the random-field disorder. 
Applying the multifractal analysis to filter the magnetisation reversal fluctuations at different scales, we distinguish the fluctuations at the beginning of the hysteresis loop from the processes at larger fields, i.e., the large-avalanches near the coercive field in the central part of the loop and the loop end.
By fixing the total number of spins $N=L_x\times L_y=10^6$ and the disorder above the critical value, we vary the domain-wall
length in different sample shapes, corresponding to the ratio
$L_x/L_y \in [1/16, 16]$.  In the following Sections, we examine the impact of these geometrical shapes to the multifractal features of the magnetisation fluctuations time series and the scale-invariance
of the avalanches in different segments of the hysteresis loop.

\section{Simulation settings with saw-tooth domain wall\label{sec:model}}
\begin{figure}[htb]
\begin{tabular}{cc} 
\resizebox{18pc}{!}{\includegraphics{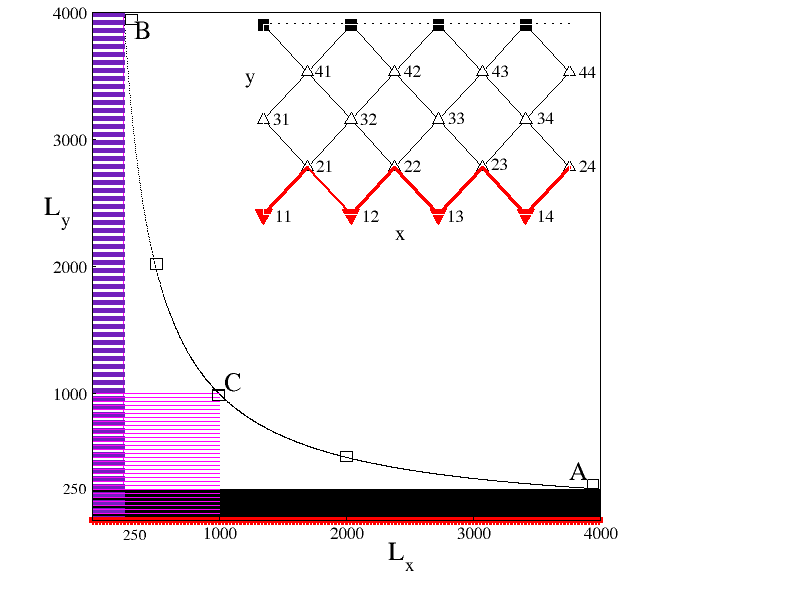}}\\
\end{tabular}
\caption{\small Inset:  Schematic presentation of the lattice with a saw-tooth domain wall (red line). The system is prepared with spin-down along the first row in $x$-direction and spin-up in the rest of sites until the top row, which is an open boundary (contains no spins). Periodic boundary conditions are applied in the transverse direction such that the left boundary in $y$-direction matches the right boundary,
  for instance, the site 31 matches 34, 44 matches 41 and so on, forming a cylindrical shape. The main figure depicts three shapes used in the simulations, termed A, B and C, which preserve the volume of $N=10^6$ spins. The domain wall is always prepared along the $L_x$ side.}
\label{fig-schema}
\end{figure}

\begin{figure*}[htb]
\begin{tabular}{cc} 
\resizebox{18pc}{!}{\includegraphics{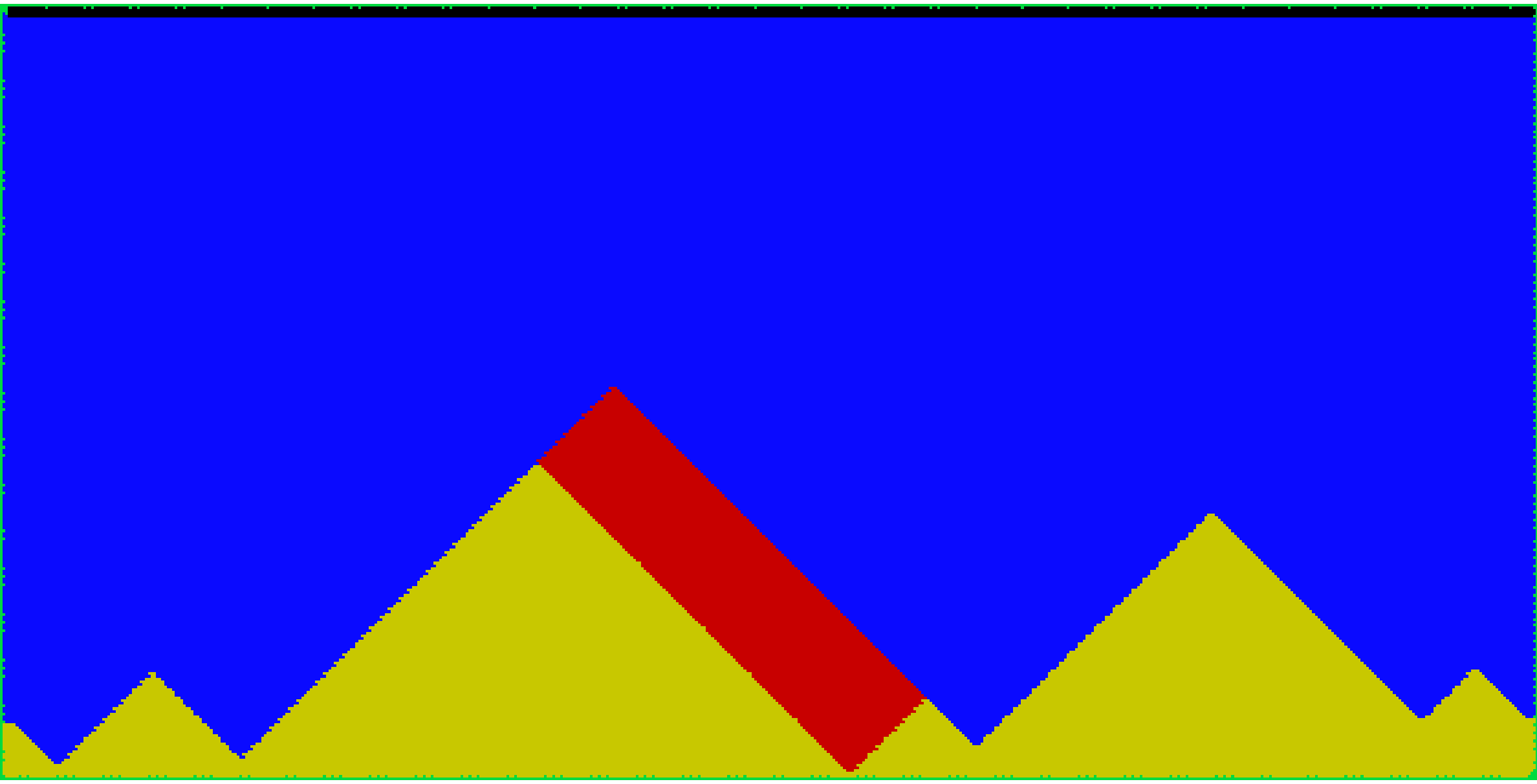}}&
\resizebox{18pc}{!}{\includegraphics{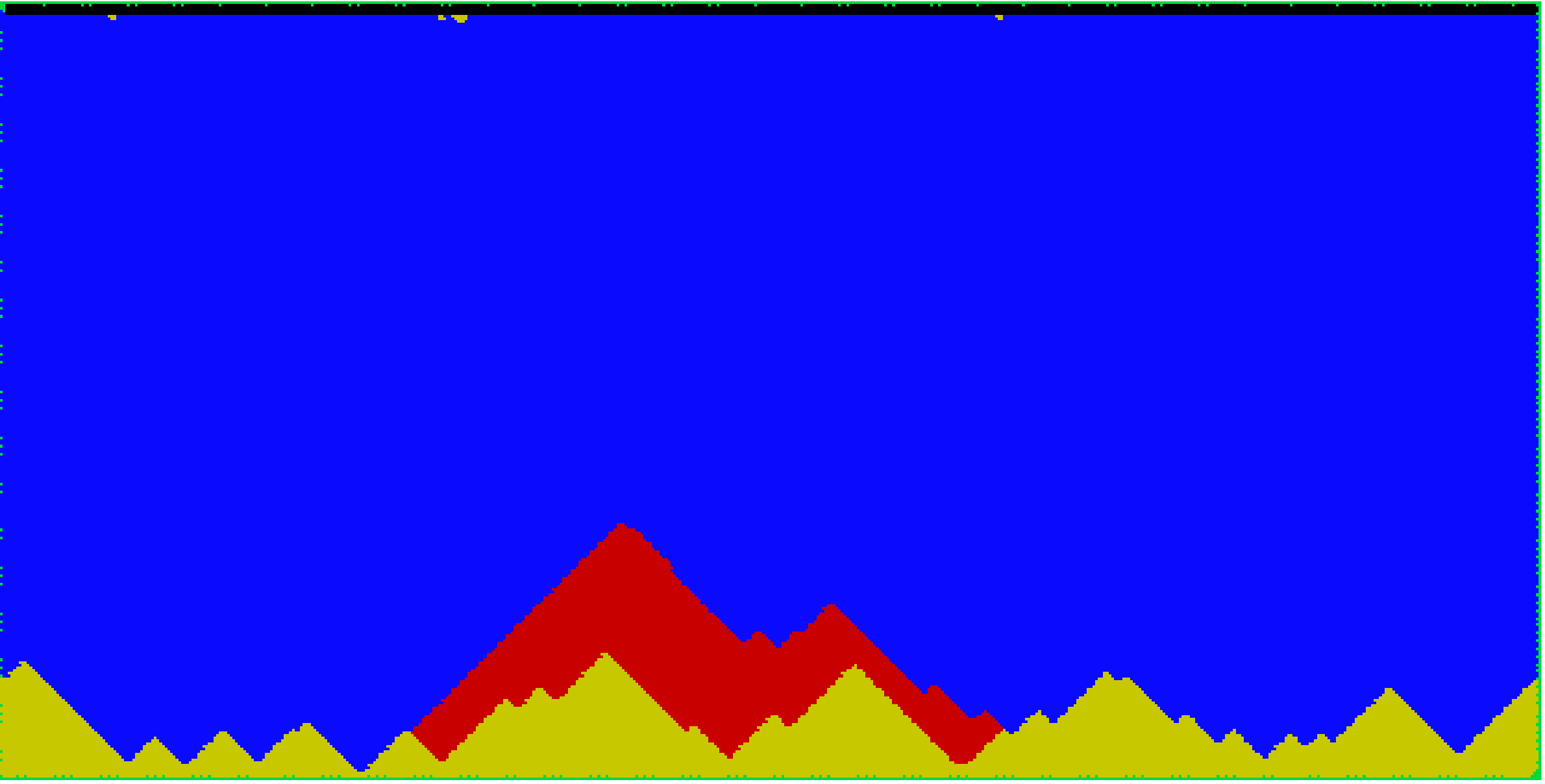}}\\
\resizebox{18pc}{!}{\includegraphics{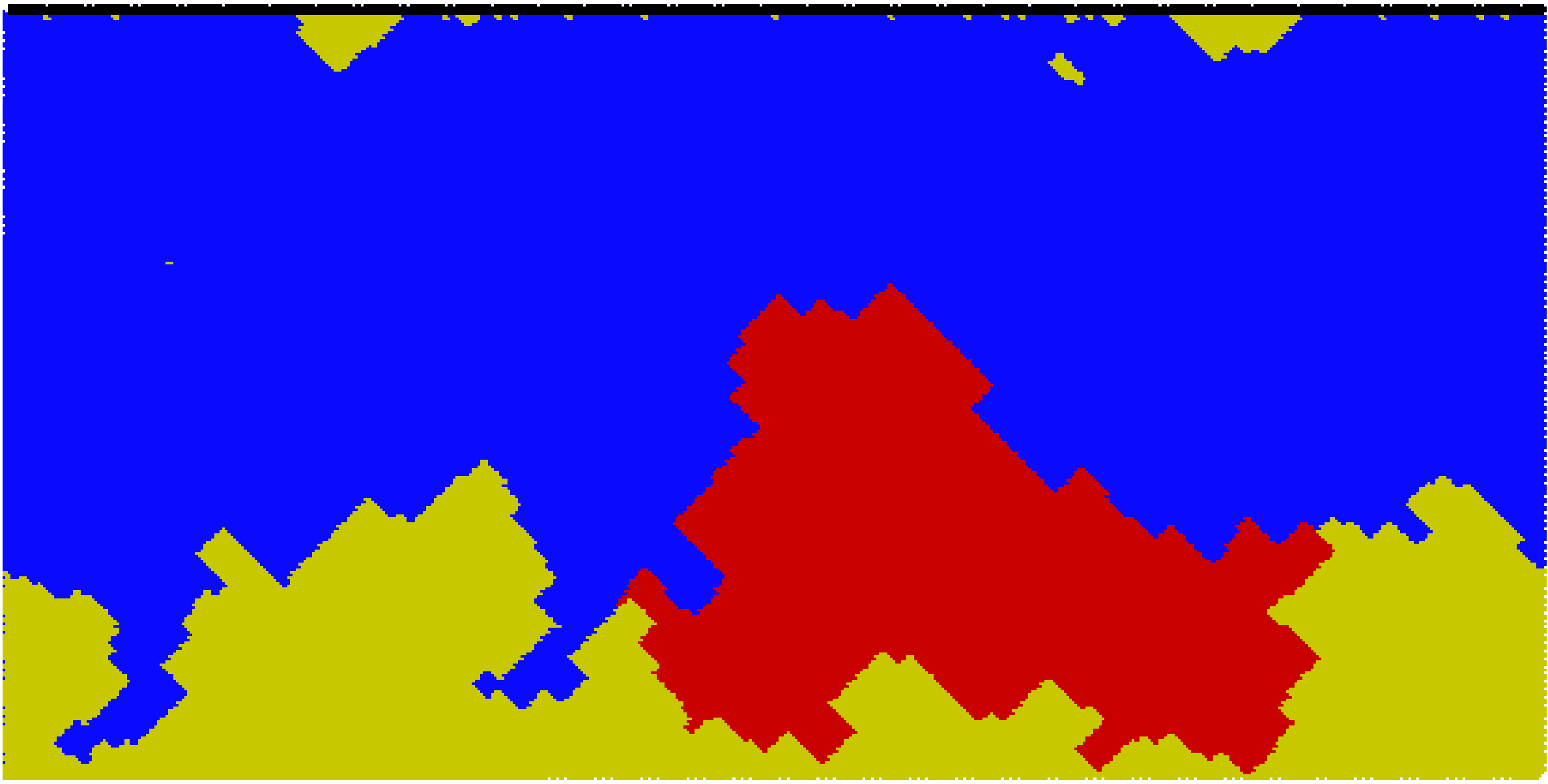}}&
\resizebox{18pc}{!}{\includegraphics{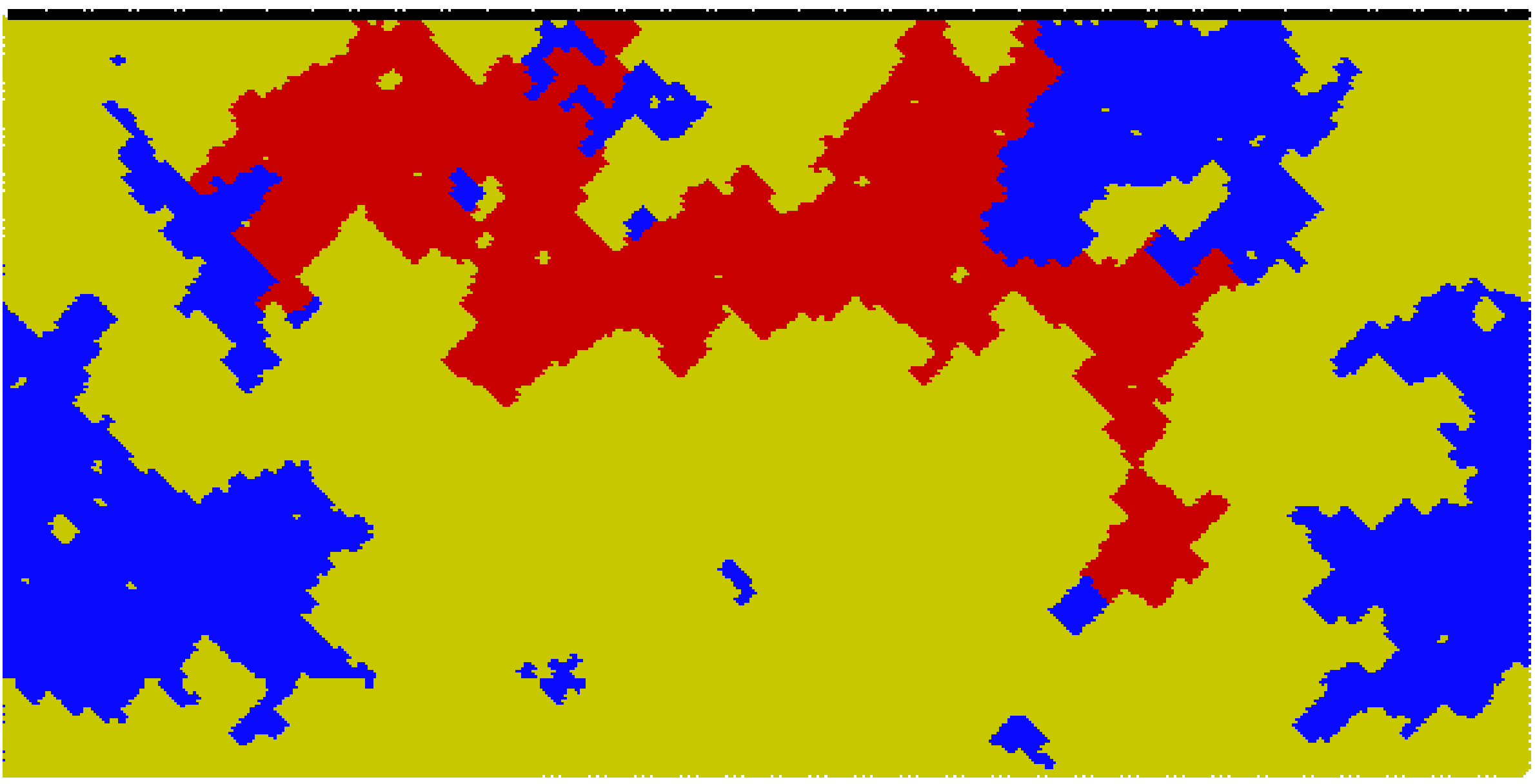}}\\
\end{tabular}
\caption{\small Snapshots of the DW position (yellow) with the most
  recent avalanche (red) and the unflipped segments of the sample
  (blue). Different pictures correspond to the disorder strength below critical  (top left), at the critical (top right) and two examples of disorder above the critical where the DW depinning is not possible, moderate (bottom left) and strong disorder (bottom right).}
\label{fig-avalanches}
\end{figure*}

The two-dimensional square lattice is rotated by 45 degrees, see Fig.\ \ref{fig-schema}, and the DW is created along the sample side in the x-direction as follows. The spins in the first row are set to $S_{1x}=-1$ while the rest of sites have $S_{ij}=+1$ excluding  the top row, which represents an open boundary with $S_{ix}=0$.  The periodic boundary conditions are applied in the $y$-direction.
The varying external field $B$ then drives the DW through the sample.   Starting from a large positive value $B_{max}$ and slowly decreasing along the descending branch of the hysteresis loop, the field is changed with a small rate $r\equiv \Delta B/J_0=10^{-3}$.  The spin dynamics is governed by the interaction Hamiltonian Eq.\ (\ref{eq-Ham}), as described below. The spin flips next to the wall represent the DW motion to a new energetically stable position; the number of spin flips between two consecutive stable positions of the DW define an avalanche, cf. snapshots in Fig.\ \ref{fig-avalanches}.  A quasi-static driving is applied, which means that the field is kept fixed during the avalanche propagation and then decreased again.  The DW annihilates at the open top boundary.

The ferromagnetic spin-spin interaction in the presence of the random-field defects
$h_i$ at lattice sites $i=1,2,\cdots N$ 
and the time-varying external field $B(t)$ is given by the Hamiltonian
\begin{equation} 
-{\cal{H}}=\sum_{i,j}J_{ij}S_i(t)S_j(t) +\sum_i h_i +B(t)  \equiv
+\sum_i h_i^{loc}(t)S_i(t) ,
\label{eq-Ham}
\end{equation}
where $J_{ij}=J_0 >0$ for all pairs of the neighbouring
spins. According to the second expression, the value of the local
field $h_i^{loc}(t)$ experienced by a spin $S_i(t)$ at the site $i$
and time $t$ depends on the actual values of the near-neighbour spins,
the quenched random field $h_i$ at location $i$ and the current value
of the external field $B(t)$. As usual in the study of  Barkhausen
avalanches by RFIM \cite{djole_PRL_critdisorder2011,djole_scaling2D,djole_spanningaval2014,eduard_FSS2003,eduard_spanning2004,BT_PRL1996,BT_PRE2000}, it suffices to consider the zero-temperature dynamics where the fields flip to align with the sign
of the local field to minimise the energy, i.e., 
\begin{equation}
S_i(t+1) = sign\left( h_i^{loc}(t)\right) .
\label{eq-si}
\end{equation}
Also, a quenched random field  $h_i$ is taken from
Gaussian distribution $h_i\in P(h,f)$ with a zero mean and the
variance $f$ (in units of $J_0^2$).  In the simulations, we perform
parallel updates of the whole system. That is, the local fields are
computed at each site and kept until all spins are updated, which
comprises one time step $t$ of the simulations. Then the fields are
calculated again, and the process is repeated until no spins are found that are not aligned with their current local fields. This step
comprises the end of an avalanche, after which the external field is
changed again, and so on.

Note that for the considered configuration, cf. Fig\ \ref{fig-schema}, in the absence of random fields, the spins at the second row can flip under an infinitesimally small energy.
Thus, in the absence of disorder, the hysteresis loop is
infinitely narrow and broadens according to the width of the distribution of the random fields. Therefore, this setting is suitable to test the actual impact of the random-field disorder to the avalanches in 2D-RFIM. 
In Fig.\ \ref{fig-avalanches}, we show snapshots of the
domain-wall position in four samples with different strength of
disorder. For a weak disorder, the DW accelerates with the changes in the external field, and eventually, DW depinning occurs in the middle of the hysteresis loop, resulting in a system-size avalanche. A precise analysis of the spanning avalanches in 2D-RFIM (but without built-in DW)  was done in \cite{djole_spanningaval2014}; the finite-size scaling analysis revealed that the spanning avalanches could occur below a critical disorder $f^0_c=0.54\pm0.06$. In the presence of the
extended DW, the finite-size scaling analysis performed in
\cite{BT_PRE2000} suggests that the spanning avalanches can happen for a bit higher values up to the critical disorder $f_c=0.64\pm 0.06$.  In the snapshots in Fig.\ \ref{fig-avalanches}, the spin flips which comprise the most recent avalanche (red areas) indicate that the presence of the extended DW induces anisotropic avalanches. Notably, the flips along the wall are energetically more favourable than the flips away from the wall, resulting in the elongated avalanche shapes even in a moderately strong disorder above the critical value. In this work, we consider the
region above the critical disorder where, besides the system-wide DW, other spanning avalanches are not possible. Thus, in all simulations, we fix the disorder by $f=0.8$, corresponding to the snapshot in the lower left panel of Fig.\ \ref{fig-avalanches}.

\begin{figure}[htb]
\begin{tabular}{cc} 
\resizebox{20pc}{!}{\includegraphics{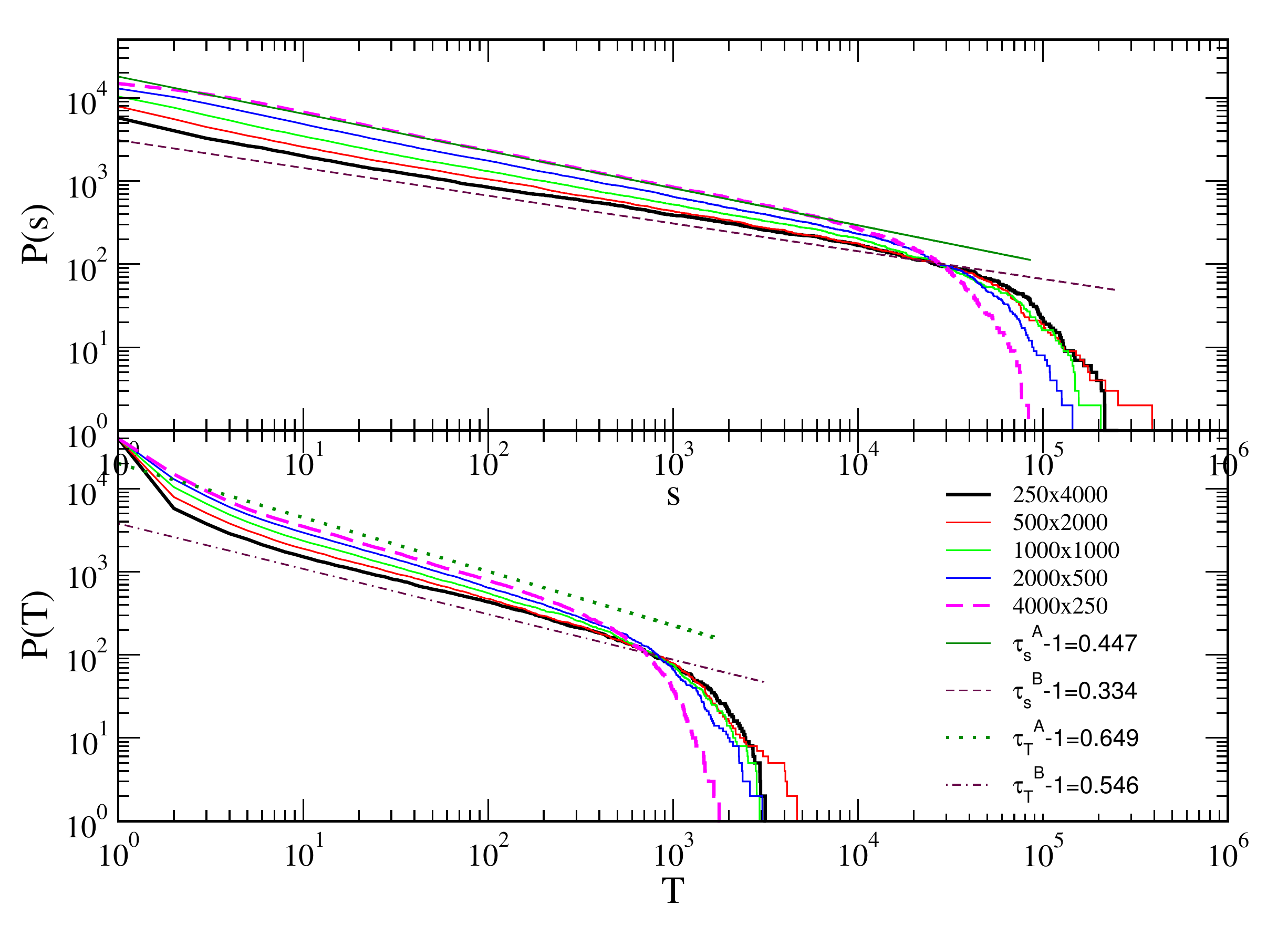}}\\
\end{tabular}
\caption{\small Cumulative distributions of the avalanche size $P(s)$ and
  duration  $P(T)$ for a fixed disorder $f=0.8$ and varied
  volume-preserving shapes, indicated in the legend and in Fig.\ \ref{fig-schema}.}
\label{fig-dspt-LxLy}
\end{figure}

As above mentioned, the presence of the saw-tooth domain wall in the configuration shown in Fig.\ \ref{fig-schema} provides a possibility to test the effects of the disorder, in particular, at the beginning of the hysteresis loop where the field is still weak and can not induce massive spin flips. Therefore the length of the DW is relevant for its motion by the same nominal disorder $f$. To explore the impact of the DW length, we fix the volume and change the sample shapes, as depicted in Fig.\ \ref{fig-schema}. The simulations are performed for five volume-preserving shapes indicated by small boxes along the hyperbolic curve in Fig.\ \ref{fig-schema}. The resulting distributions of the avalanche sizes and durations, integrated along the whole hysteresis branch,  averaged over $10\times$ realisations of random fields, are shown in Fig.\ \ref{fig-dspt-LxLy}. 
Notably, the sample shape affects the avalanches of magnetisation
reversal. In particular, the scaling exponents are higher in the stripe with the long DW (sample-A) than in the samples with gradually shorter DW, see Table\ \ref{tab-exponents}. Eventually, in the stripe with the short DW (sample-B), we recover the exponents that are numerically close to those previously found  \cite{djole_scaling2D} in the 2D-RFIM without a built-in DW. It is important to notice that the cut-offs of the distributions also vary with the DW
length, suggesting that the long DW experiences different pinning effects  than the short DW at the same nominal disorder. Moreover, these cut-offs differ from the usual exponential cutoffs.

 Our analysis in the following two sections reveals that the origin of these avalanching behaviours is in the altered dynamics of the DW in different sample shapes. Following the dynamical analysis, we will show that various segments of the hysteresis loop contribute
differently to the overall avalanching dynamics, depending on the length of the DW.

\section{Dynamics of the magnetisation reversal in different segments
  of the hysteresis loop\label{sec:dynamics}}
\begin{figure}[htb]
\begin{tabular}{cc} 
\resizebox{18pc}{!}{\includegraphics{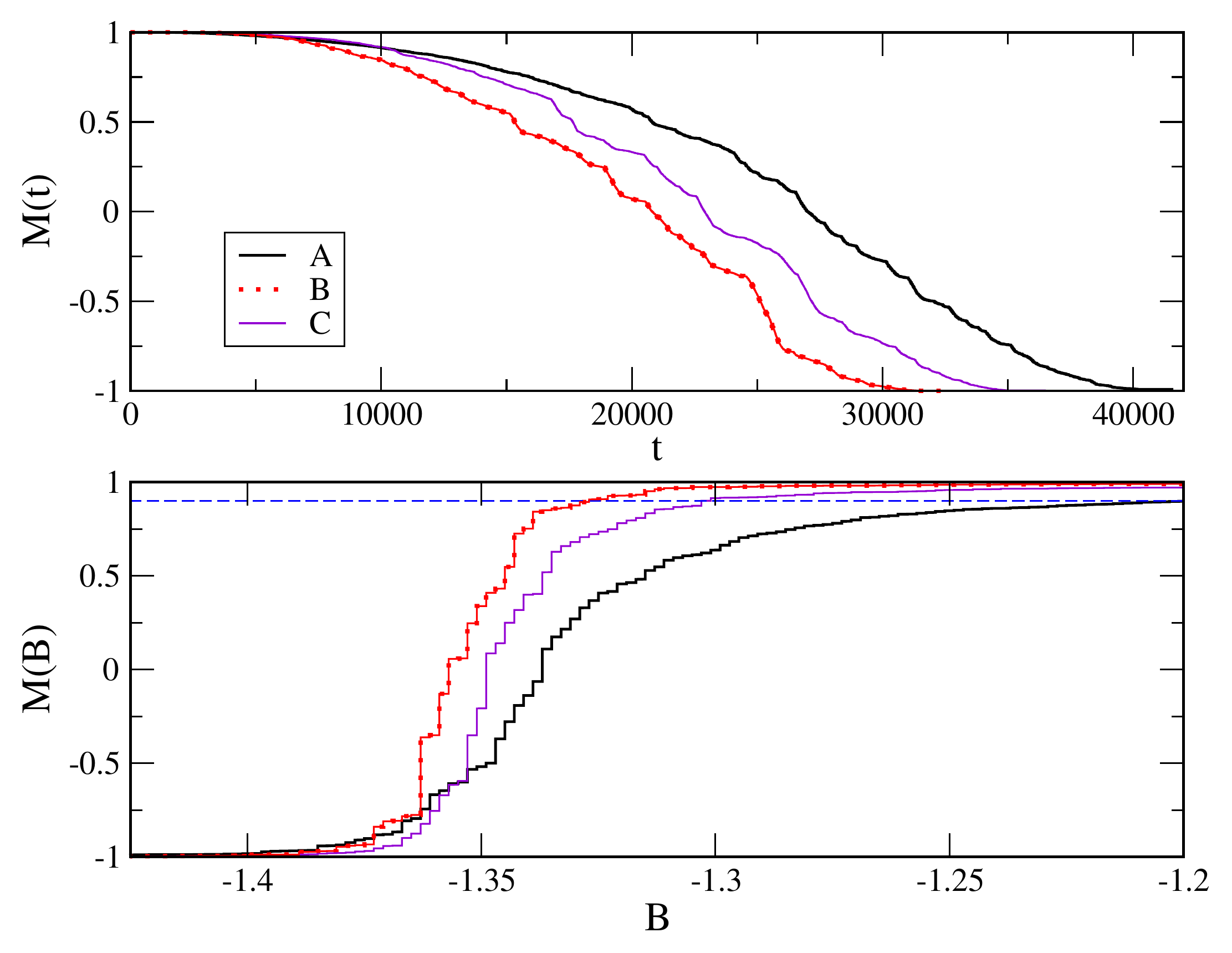}}\\
\end{tabular}
\caption{\small Magnetisation as function of time (top) and the
  external field (bottom) for three samples marked as (A), (B)
  and (C), cf.\  Fig.\ \ref{fig-schema}.}
\label{fig-H-M-t}
\end{figure}

Following the statistics of the avalanches, mentioned above, a detailed inspection of the magnetisation reversal process from +1 to -1 shows the different course of the process in samples with different shapes, i.e., the length of  DW. Here, we examine and compare these processes in three prototypal shapes indicated as A, B and C in Fig.\ \ref{fig-schema}. In particular, these are two narrow stripes (2D nano-wires) with the DW set along the longer side (sample A) and the shorter side (sample B), and the corresponding symmetrical shape with the same number of spins (sample C). The top panel of Fig.\ \ref{fig-H-M-t} shows how the magnetisation evolves with time in these samples. In particular, the reversal occurs in the shortest time in sample B having the short DW, while the evolution is slower in the shape A with the long DW, and the case of the symmetrical shape C is in between these two. Analogously, the hysteresis loop in the lower panel of Fig.\ \ref{fig-H-M-t} receives different forms; here, the long DW in the sample A results in the most slanted loop while the symmetrical and short DW cases lead to a more rectangular loop. Notice that the shape of the hysteresis loop near the coercive field, where the magnetisation eventually changes the sign, comprises of some large jumps (avalanches) of the magnetisation. While these large but finite jumps are in accord with the pinning by the supercritical disorder, they appear to vary with the sample shape, that is with lengths of the DW.

\begin{figure}[htb]
\begin{tabular}{cc} 
\resizebox{21.8pc}{!}{\includegraphics{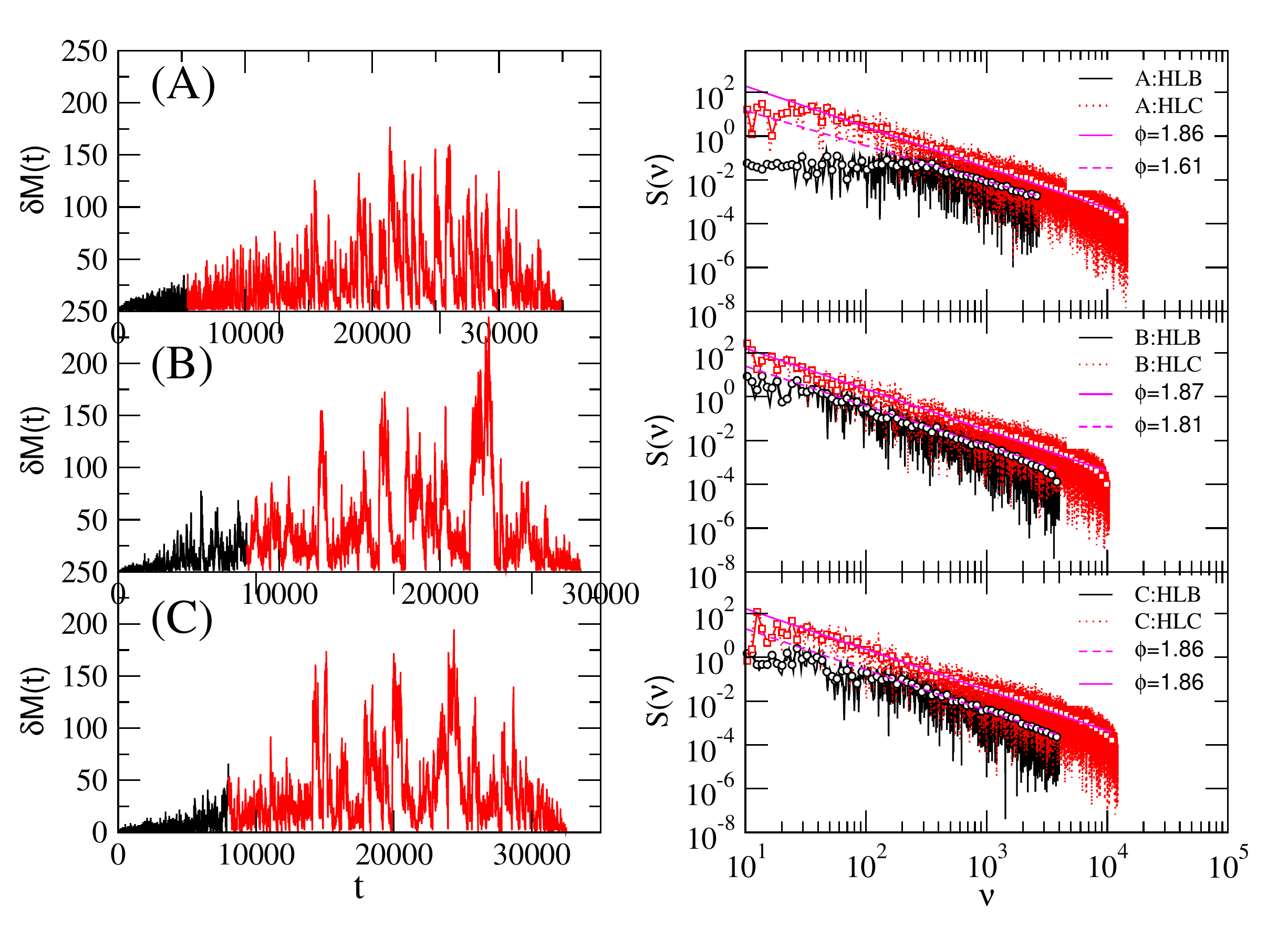}}\\
\end{tabular}
\caption{\small The magnetisation fluctuation time series (left) and
  their power spectra (right) in samples marked as (A), (B)
  and (C), cf.\ Fig.\ \ref{fig-schema}. Black and red colour mark
   HLB and HLC segments of the hysteresis loop, respectively.}
\label{fig-nt-ABC}
\end{figure}
Moreover, the fluctuations of the magnetisation along the hysteresis loop also differ with the varied shapes of the sample. For the
samples A, B and C, in Fig.\ \ref{fig-nt-ABC}, we show the time
series of the magnetisation changes $\delta M(t)$, which represents the number of reversed spins per time step $t$. In this type of signal, known as Barkhausen noise, an avalanche can be seen as a part of the signal above the baseline (level zero in this case).  In agreement with the shape of the hysteresis loop
in Fig.\ \ref{fig-H-M-t}b, some large avalanches occur in the
central part of the loop. They are most pronounced in sample B,
which also features the fastest reversal. The large avalanches with
gradually smaller sizes also occur in the central part of the loop in sample C and A, compatible with the extension of the respective reversal times.
According to Fig.\ \ref{fig-H-M-t}, the major differences occur at the
beginning of the loop in different DW morphologies. Therefore, we
consider the initial part of the signal, comprising of 5\% of the
total magnetisation reversal,  as the hysteresis-loop-beginning (HLB) process, to be distinguished from the remaining signal, which is termed hysteresis-loop-central (HLC) process. In the following, we
separately analyse the fluctuation features of these segments of the signal. 

In the right panels of Fig.\ \ref{fig-nt-ABC}, we show the power spectra of these segments of the signal for sample shapes A, B and C.  
Notably, all signals exhibit temporal correlations, which are manifested in the power-law decay of the power spectrum as $S(\nu)\sim \nu^{-\phi}$ for a range of high frequencies $\nu$. The observed values of the exponent  $\phi=1.61 \pm 0.12$ in the HLB segment of sample A is the one that differs from $\phi=1.84 \pm 0.06$ in the HLC. In the other two samples, the exponents are similar, and all are found in the range $1.80-1.90$, probably dominated by the increasing trend of the signal with the increased field along the loop. 

Further, we investigate the potential differences in these
fluctuations at a wide range of scales, which can be filtered by multifractal analysis of the corresponding time series.

\begin{figure}[htb]
\begin{tabular}{cc} 
\resizebox{18pc}{!}{\includegraphics{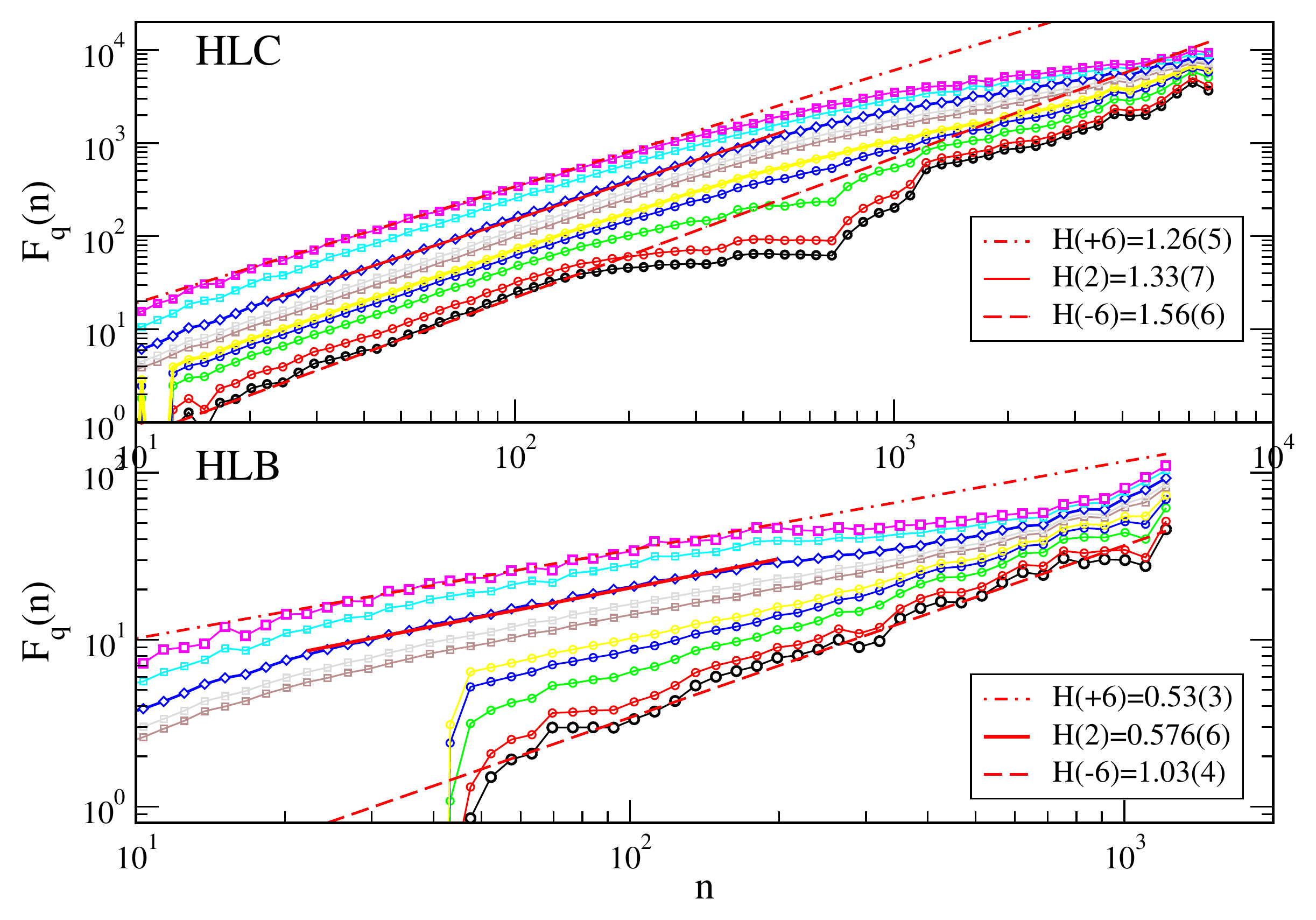}}\\
\resizebox{18pc}{!}{\includegraphics{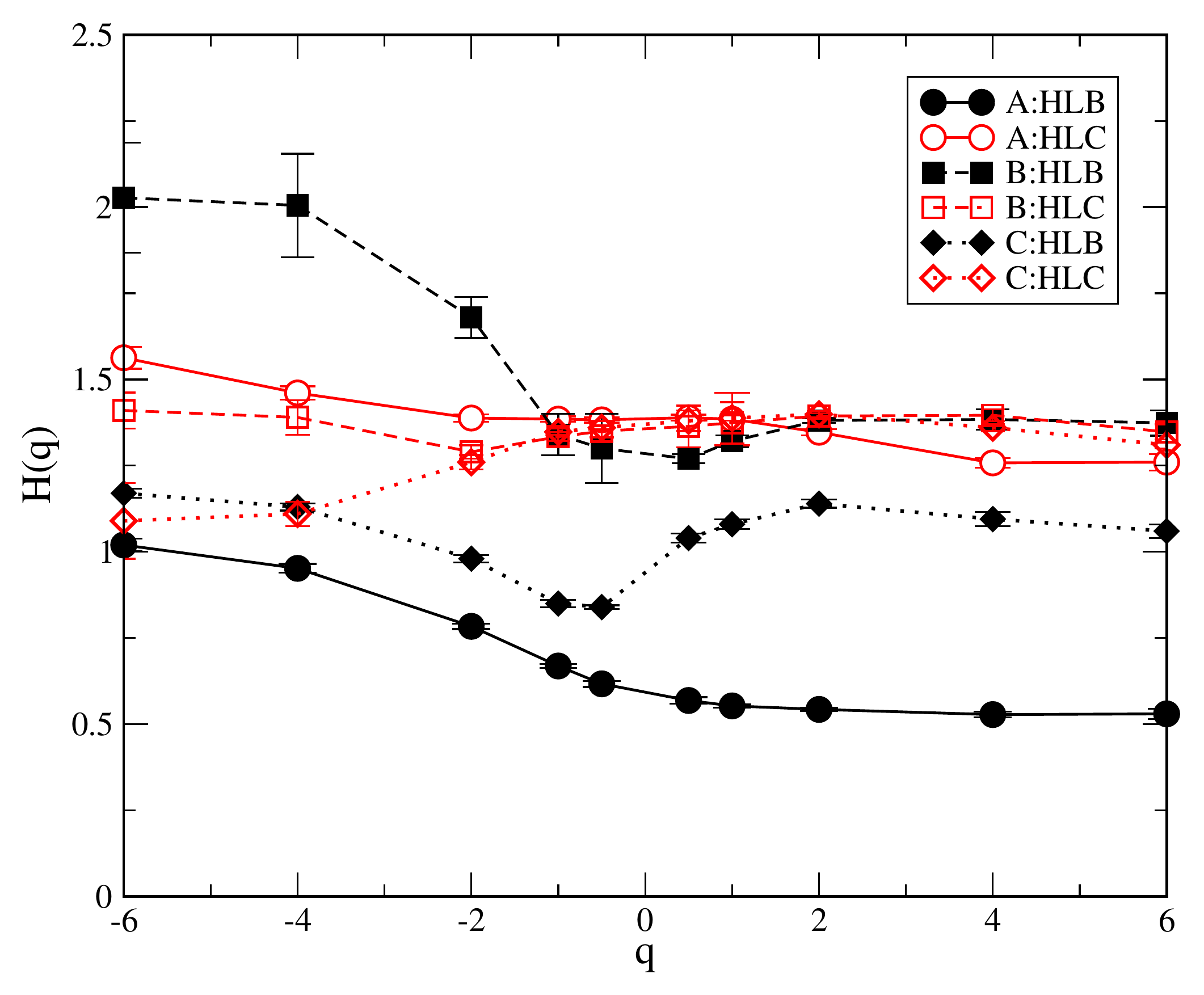}}\\
\end{tabular}
\caption{\small Top panels:  Fluctuation function $F_q(n)$ vs. interval length $n$
  for $q\in [-6,+6]$ bottom to top curves; two panels correspond to
  the signal originating from 
   HLC and HLB segments of the hysteresis loop  for sample A. Bottom panel: The generalised Hurst
  exponents $H(q)$ plotted against $q$ for  HLB and HLC fluctuations
  in all three samples A, B and C, as indicated in the legend.}
\label{fig-mfra}
\end{figure}

We apply the detrended multifractal analysis
\cite{MFRA-uspekhi2007,DMFRA2002,dmfra-sunspot2006,dmfra_drozdz2015,BT_MFRbhn2016} of time series. The procedure  utilises the underlying self-similarity, which is suitable for the analysis of Barkhausen noise
signals \cite{BT_MFRbhn2016}; it aims at exploring the scale
invariance of the $q-$th order fluctuation function $F_q(n)$, defined
below, which is expressed by a range of scaling exponents $H(q)$
(generalised Hurst exponent). Here, $q$ is a real number taking a broad range of positive and negative values by which different segments of a multifractal time series get amplified such that they become self-similar to the rest of the signal. The standard procedure of the analysis consists of several steps
\cite{DMFRA2002,dmfra-sunspot2006,dmfra_drozdz2015,BT_MFRbhn2016}, in particular:

First, the profile of the time series $\delta M(k)$, $k=1,2,\cdots
T_{max}$ is the time step index,  is obtained by integration 
\begin{equation}
Y(i) =\sum_{k=1}^i(\delta M(k)-\langle \delta M\rangle)  \ .
\label{eq-profile} 
\end{equation}
and divided into non-overlapping segments of equal length $n$.  The
process is repeated starting from the end of the time series; thus in
total $2N_s=2Int(T_{max}/n)$ segments are considered for each time series.
  Then, the  local trend $y_\mu(i)$ is determined  (by non-linear
  interpolation) at each segment $\mu=1,2\cdots N_s$, and the standard
  deviation around  the local trend is computed:
\begin{equation}
 F^2(\mu,n) = \frac{1}{n}\sum_{i=1}^n[Y((\mu-1)n+i)-y_\mu(i)]^2  \ ,
\label{eq-F2}
\end{equation}
and similarly, $F^2(\mu,n) =
\frac{1}{n}\sum_{i=1}^n[Y(N-(\mu-N_s)n+i)-y_\mu(i)]^2$ for $\mu
=N_s+1,\cdots 2N_s$. The $q$-th order fluctuation function $F_q(n)$ is
then obtained  for the segment length $n$ and averaged over all
segments. 
\begin{equation}
F_q(n)=\left\{(1/2N_s)\sum_{\mu=1}^{2N_s} \left[F^2(\mu,n)\right]^{q/2}\right\}^{1/q} \sim n^{H(q)}  \ .
\label{eq-FqHq}
\end{equation}
For a given $q$, the fluctuation function is plotted against varied segment length $n$ and a scaling region (a straight line on double-log plot) examined to extract the corresponding Hurst exponent $H(q)$.
The considered segment lengths vary in the range $n\in[2,int(T_{max}/4)]$.
The distortion parameter $q$ takes a range of real values, here $q\in
[-6,+6]$, that allows to amplify  \textit{small fluctuation segments} by the negative values of $q$, and the segments with \textit{large
  fluctuations},  for the positive values of $q$. Note that for $q=2$, we obtain the well-known Hurst exponents of the standard deviation of the time series. Hence, if the time series is a monofractal, all values of the generalised Hurst exponent coincide and are equal (within the numerical error bars) to the standard Hurst exponent. In contrast, the multifractal time series exhibit different scaling exponents, that is, $H(q)$ is the function of $q$ at least in a certain range of scales $n$, see the examples in Fig.\ \ref{fig-mfra}. As discussed in the literature
\cite{MFRA-uspekhi2007,DMFRA2002,dmfra-sunspot2006,dmfra_drozdz2015}, the $H(q)$ dependence on $q$ can be mapped to the singularity spectrum$\Psi(\alpha)$
of the time series via Legendre transform  of $\tau (q)$, i.e.,
$\Psi(\alpha)=q\alpha -\tau (q)$, where $\alpha =d\tau/dq =
H(q)+qdH/dq$. Here,  $\tau(q)$ is the corresponding scaling exponent
associated with the standard measure (box probability) in the
partition function method. 

As above mentioned, here we consider six different time series corresponding to the HLB and HLC in each of the three sample shapes A, B and C. 
In Fig.\ \ref{fig-mfra}, we show the example of the
fluctuation functions for the HLB and HLC part of the Barkhausen signal for sample A. This Figure shows that the fluctuation
originating from the HLB scale differently than the fluctuations in
the HLC. Appart from the wider multifractal spectrum in HLB, the
values of the exponents $H(q)$ remain in the range below one, see the bottom panel in Fig.\ \ref{fig-mfra}. 
This spectrum suggests that the profile is a fractional Brownian motion and, consequently, the original signal $\delta M(t)$, in this case, is a fractional Gaussian noise. In contrast, the HLC fluctuations result with a more narrow spectrum with the exponents $H(q)>1$, which is compatible with the sum of the fractional Brownian signals
\cite{dmfra-sunspot2006,BT_MFRbhn2016}. The corresponding scaling exponents $H(q)$ for all considered sample shapes are shown in the bottom panel in Fig.\ \ref{fig-mfra}. Noticeably, the enhanced small fluctuations (negative $q$-region of the spectra) are observed for HLB in all three sample shapes. The difference is that, for the sample B comprising of a short DW, the entire spectrum remains above the line $H=1$. Whereas, in sample C with the symmetrical shape, the profile $H(q)$  for HLB fluctuations contains a mixture of the spectra seen in the samples A and B. 
In contrast, the spectra of the fluctuations in the HLC of all sample shapes are similar, cf. empty symbols in Fig.\ \ref{fig-mfra}.  We discuss these findings  in the final Section\ \ref{sec:conclusions}. In the following Section, we analyse these signals at a mesoscopic scale of avalanches occurring in the samples of different shapes.

\section{Statistics of the magnetisation reversal avalanches in
  different sample shapes \label{sec:statistics}}
\begin{figure}[htb]
\begin{tabular}{cc} 
\resizebox{18pc}{!}{\includegraphics{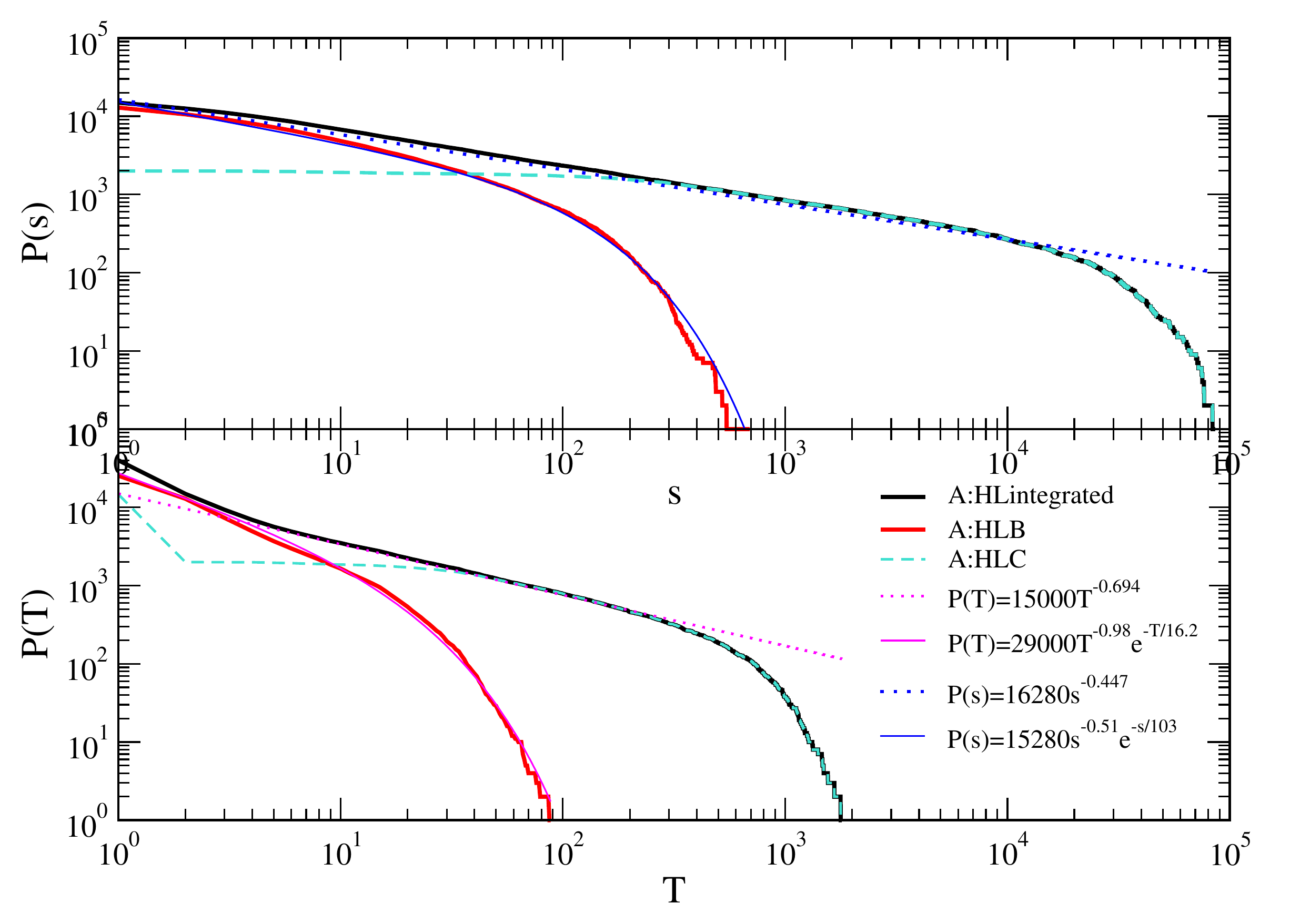}}\\
\end{tabular}
\caption{\small Cumulative distributions of the avalanche size $P(s)$ and
  duration $P(T)$ for HLB  (red line) and  HLC segment (dashed magenta
  line) and the distribution integrated along the hysteresis branch
  (black line); the results are for the narrow stripe $L_x=16L_y$ with a long DW (sample A). }
\label{fig-dspt-A}
\end{figure}

\begin{figure}[htb]
\begin{tabular}{cc} 
\resizebox{18pc}{!}{\includegraphics{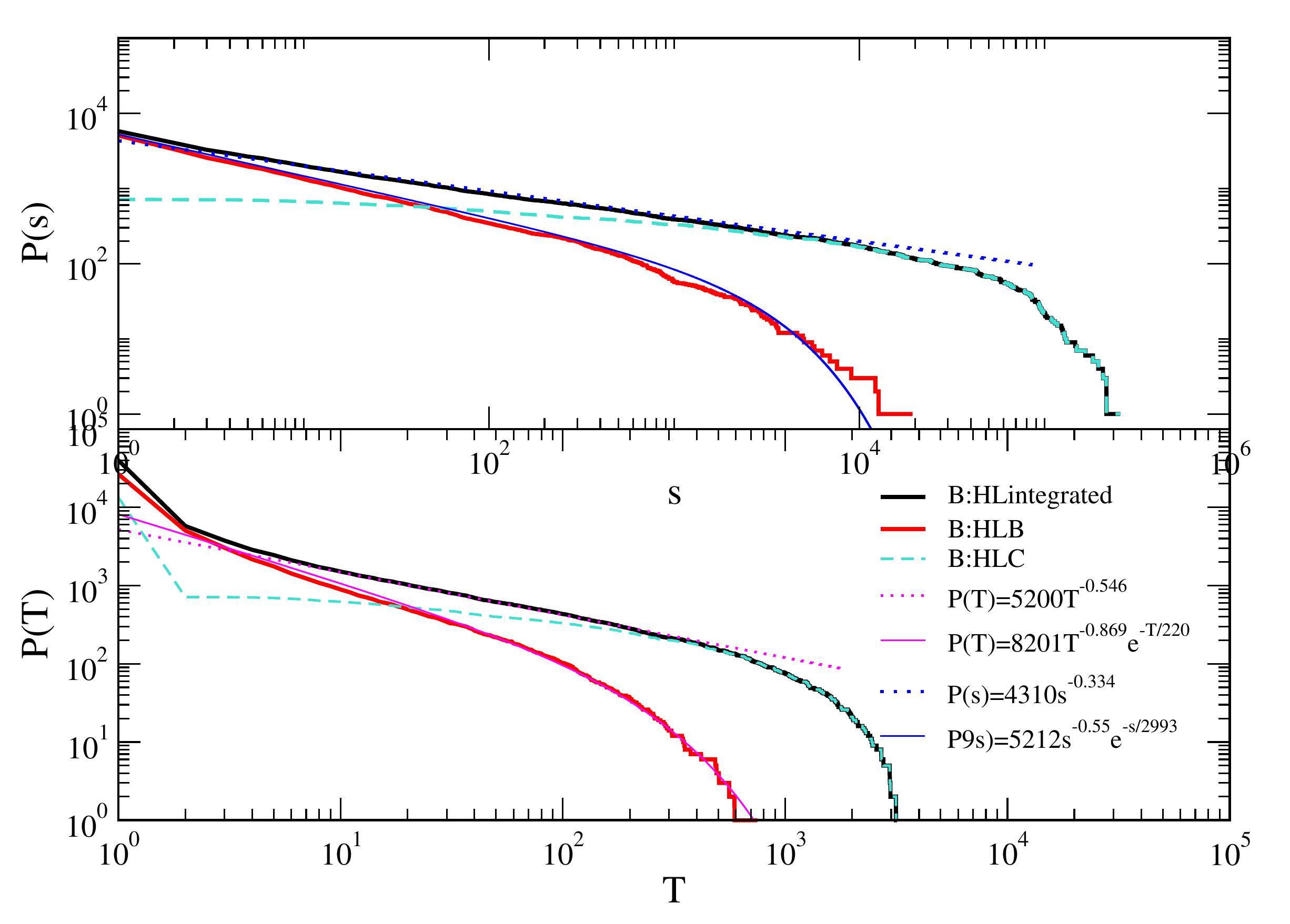}}\\
\end{tabular}
\caption{\small Cumulative distributions of the avalanche size $P(s)$ and
  duration $P(T)$ for HLB  (red line) and  HLC segment (dashed magenta
  line) and the distribution integrated along the hysteresis branch
  (black line); the results are for the
  narrow stripe  $L_x=L_y/16$  with a short DW (sample B).}
\label{fig-dspt-B}
\end{figure}

\begin{figure}[htb]
\begin{tabular}{cc} 
\resizebox{18pc}{!}{\includegraphics{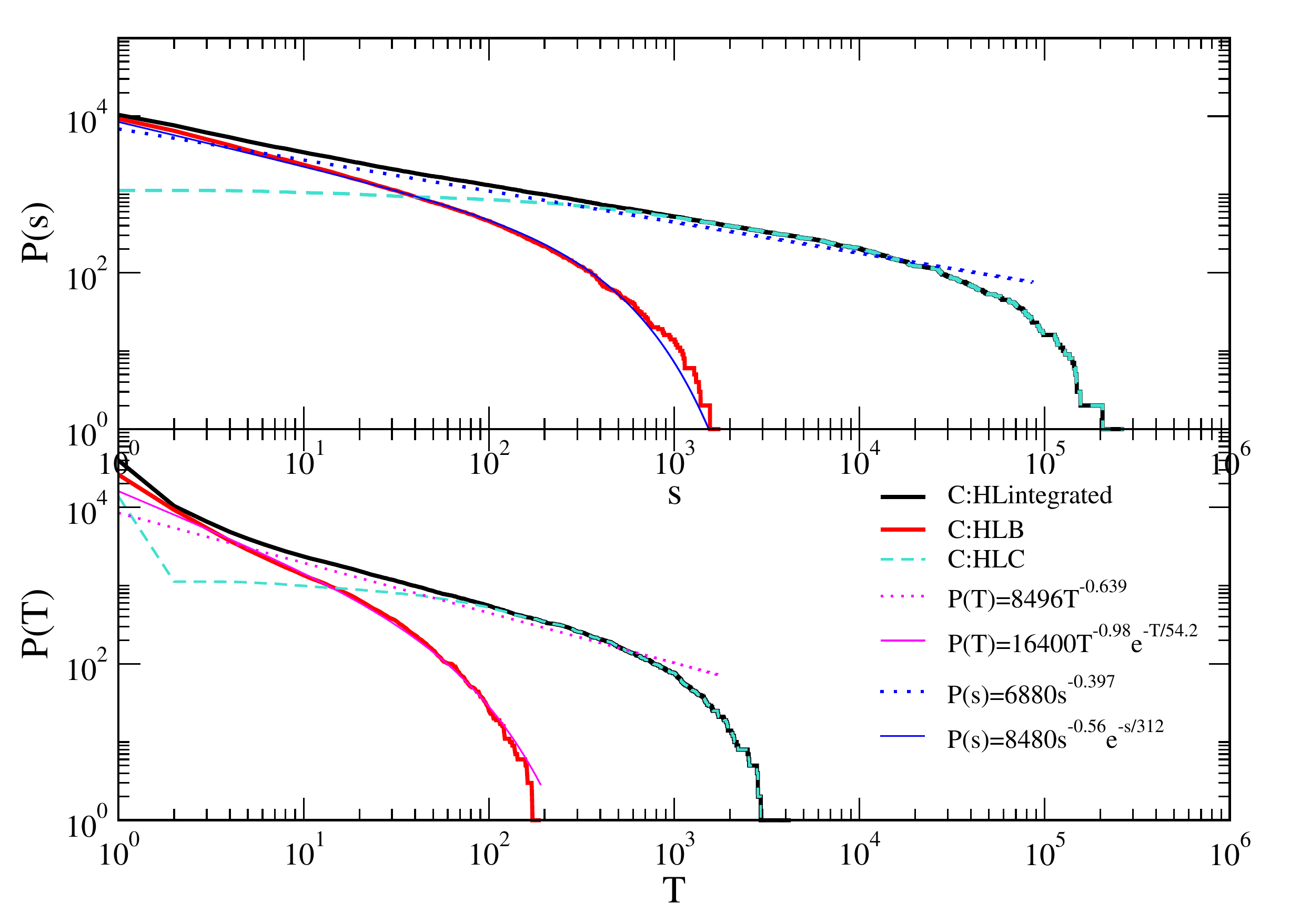}}\\
\end{tabular}
\caption{\small Cumulative distributions of the avalanche size $P(s)$ and
   duration $P(T)$ for HLB  (red line) and  HLC segment (dashed magenta
  line) and the distribution integrated along the hysteresis branch
  (black line); the results are for the symmetrical sample $L_x=L_y$ (sample C).}
\label{fig-dspt-C}
\end{figure}

By definition, an avalanche represents a clustering of events along the time
series; hence, it consists of a number of the elementary pulses. 
According to the dynamical analysis of the magnetisation
fluctuations time series in Sec.\ \ref{sec:dynamics}, we expect that the observed different
stochastic processes have an impact to the avalanches. 
Specifically, we separate the contribution arising in the process at
the beginning of the hysteresis loop, where the disorder is essential
to the DW motion,  from the processes in its centre, where the field
is strong enough to induce maximal avalanches in a given DW
configuration and the strength of disorder. From the simulations, we determine the avalanche
distributions associated with the HLB and HLC for the considered sample
shapes. The distributions are averaged over 10 random-field
configurations. For each sample shape, the corresponding
configurations of the random fields are generated starting from the same seed. 
The results depicted in Figs.\ \ref{fig-dspt-A}, Fig.\ \ref{fig-dspt-B} and
Fig.\ \ref{fig-dspt-C} are for the samples A,B and C, respectively.

In sample A, featuring the long DW, we find that the distribution of
size $P(s)$ and duration $P(T)$ of the avalanches recorded in the
initial part of the hysteresis loop can be fitted by power-law decay
up to the exponential cut-off. Specifically, we have
\begin{equation}
P(T)=aT^{-(\tau_T-1)}e^{-T/T_0} \ ,
\label{eq-plexp}
\end{equation} 
for the avalanche duration $T$, and similarly for the avalanche size $s$, cf. legend in Fig.\ \ref{fig-dspt-A}. 
The fitted values for the corresponding exponents
$\tau_s-1=0.51\pm 0.04$, $\tau_T-1=0.98\pm 0.05$ indicate a familiar mean-field universality class for the HLB criticality. In contrast, the distributions integrated over the entire branch of the hysteresis loop exhibit a scale-invariance with different exponents and non-exponential cutoffs. In particular, the exponents are $\tau_s-1=0.447$ and $\tau_T-1=0.694$, within the numerical error-bars $\pm 0.009$, suggesting a new class of the avalanching dynamics. Importantly, in the scaling and cutoff region, these distributions coincide with the distributions obtained from the central part of the hysteresis loop (dashed line). Hence, the avalanching
dynamics accompanying the motion of the long DW in the central part of the hysteresis loop is responsible for the new critical behaviour. A more detailed discussion is given in Section\ \ref{sec:conclusions}.

The presence of the short DW in sample B manifests in somewhat
different avalanche behaviour, as shown in Fig.\ \ref{fig-dspt-B}.  In
particular, for the HLB, fits of the distributions by the expression
(\ref{eq-plexp}) lead to the exponents $\tau_s=0.551$ and
$\tau_T-1=0.869$, numerically close to the exponents obtained in
\cite{djole_scaling2D} for the 2D-RFIM without any built-in DW.  The
loop-integrated distributions, in this case, lead to the exponents $\tau_s-1=0.33$ and $\tau_T-1=0.54$, deviating from the distribution in HLC, which are almost flat (see below). 
It is interesting to note that these numerical values are close to the
exact values $\tau_s=4/3$ and $\tau_T=3/2$ derived for the avalanches in the directed sandpile automata \cite{DD1990},  representing the probability of a meeting of two
random walks initiated at the same point. 
The simulations in the symmetrical sample, where the DW length is equal to the transverse distance $L_y$, confirm that the change of the exponents is indeed caused by the extension of the DW. The results for sample C are shown in Fig.\ \ref{fig-dspt-C},
where the fits with the expression (\ref{eq-plexp}) give the mean-field exponents for HLB avalanches, and 
the corresponding exponents in the HLC and the loop-integrated distributions are in between the exponents in samples A and B.

\begin{figure}[htb]
\begin{tabular}{cc} 
\resizebox{18pc}{!}{\includegraphics{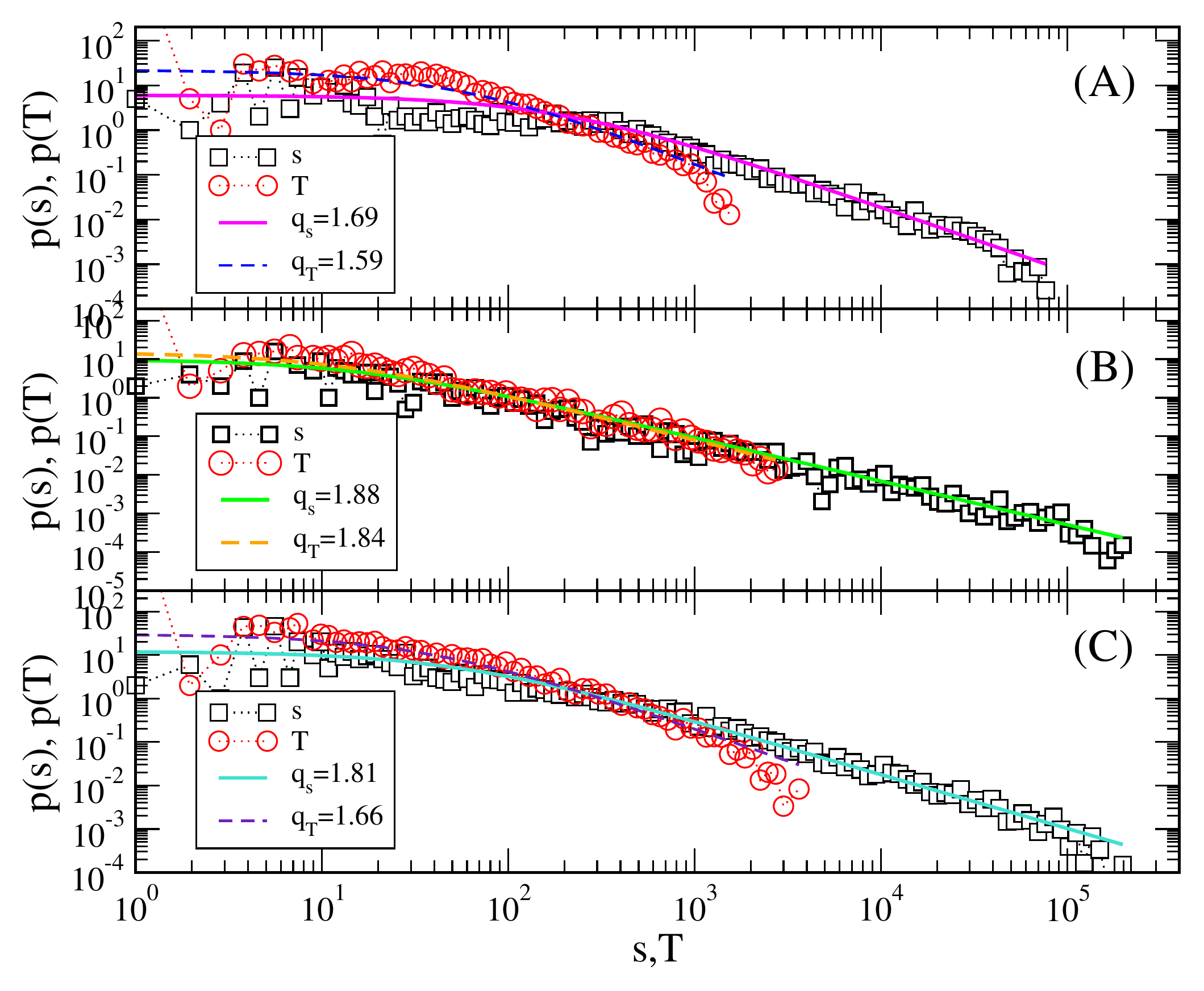}}\\
\end{tabular}
\caption{\small Differential distributions of the avalanche size $s$  and duration $T$ in the central part of the hysteresis loop for three sample shapes A, B and C (data are log-binned). The fits according to Tsallis density
function Eq.\ (\ref{eq-qexp}) with the indicated $q_s$ and $q_T$ values.}
\label{fig-HLC-Tsallis}
\end{figure}

To further explore the contribution of the dynamics from the central segments of the hysteresis loop to the occurrence of the new universality classes of avalanches of the DW slips, we determine and fit the \textit{probability density} of the size and duration of these avalanches.  The distributions for all sample shapes, averaged over $20\times$ realisations of random fields,  are shown in Fig.\ \ref{fig-HLC-Tsallis}.

In contrast to the expression (\ref{eq-plexp}) that applies to the
avalanche distributions in HLB, the differential distributions for the
avalanche size $s$ in HLC can be better fitted by Tsallis density function \cite{tsallis-book2004,pavlos_q-returns2014}
\begin{equation}
p(s)=B_s\left[1-(1-q_s)s/s_0\right]^{\frac{1}{1-q_s}}  \ ,
\label{eq-qexp}
\end{equation}
and similarly for the distribution of the duration $T$. For varied sample shapes, these distributions appear to have different slopes, i.e., nonextensivity parameters $q_s$ and $q_T$, as indicated in the figure legends. 
 Specifically, with the decrease of the ratio $L_x/L_y$ the scaling
 exponent $\tau_s=1/(q_s-1)$ systematically decreases  from
 1.44 (in sample A) to 1.24 (sample C) to 1.13 (sample B). 
Hence, the exponent of the corresponding cumulative HLC distribution in Fig.\ \ref{fig-dspt-B} is close to zero within the numerical error bars (fitting the slope for sample B in Fig.\ \ref{fig-HLC-Tsallis} gives $\tau_s=1.08\pm 0.09$).
It is interesting to compare these geometrical effects on the DW-slip
avalanches with the experimentally observed tuning of the scaling
exponent $\tau_s$ in Ref.\ \cite{koreans_JApplPhys2008}.
Considering the distribution of the duration of avalanches, we also
find that the exponent $\tau_t=1/(q_t-1)$ varies from 1.69 (sample A)
to 1.51 (sample C) and 1.19 (sample B).
A possible origin of Tsallis distribution in the context of the
domain-wall avalanches, and the observed different classes of
criticality  are also discussed in Section\ \ref{sec:conclusions}.

\section{Discussion and Conclusions\label{sec:conclusions}}

We have considered magnetisation reversal processes in a model of two-dimensional ferromagnet with the morphology that allows built-in saw-tooth DW of the length $L_x$ along one side of the sample.  By fixing the number of spins, we considered several sample shapes by varying the ratio $L_x/L_y$.
Given the periodic boundaries along $L_y$ boundary, the considered samples comprise of cylindrical shapes with a different diameter $L_x$.  Simulating zero-temperature RFIM dynamics under slow field ramping, we have analysed the motion of the domain wall through the sample until it is annihilated at the open top boundary, comprising the complete magnetisation reversal along the hysteresis branch. 
Our analysis in this paper focuses on the fluctuations of magnetisation at different scales, which are filtered by the detrended multifractal analysis of the Barkhausen noise time series, and their impact to the avalanches of the domain-wall slips. Also, we fixed the strength of the random-field disorder $f=0.8>f_c$ that prevents the domain-wall depinning or the appearance of other (linear) spanning avalanches in the process. In the considered morphology, the avalanches always start from the domain-wall position (initially at the bottom boundary of the sample).

Our main findings are that different sample shapes (allowing different lengths of the domain wall) have implications for the magnetisation reversal noise at all scales from the elementary pulses to avalanches of the domain-wall slips. Moreover, the impact of the domain-wall length on its dynamics differs at the beginning of the hysteresis loop (HLB), where the disorder effects are dominant, from the remaining part of the
hysteresis loop (HLC), corresponding to the stronger external field.
Specifically, we find that
\begin{itemize}
\item the magnetisation fluctuations at the beginning of the hysteresis loop exhibit multifractal features with a broad spectrum; enhanced small changes occur in all sample shapes. However, the multifractal spectrum gradually shifts from the region of the fractional Gaussian noise (for large DW, sample A) to fractional Brownian motion fBM (in the case of short DW, sample B). For the processes with large avalanches in HLC,  a narrow spectrum corresponding to fBM is found, which is less sensitive to the DW length.
\item the scale-invariance of the avalanches in HLB belongs to the
familiar mean-field universality class, in particular for a long DW. In sample with the short DW, the exponents are numerically close
  to those in 2D-RFIM without a built-in DW \cite{djole_scaling2D}.
\item  The avalanche distributions in HLB  obey a commonly known power-law decay with the exponential cutoffs. In contrast, the avalanches in HLC follow Tsallis density distribution with the power-law tails. The corresponding scaling exponents are tuned by changing the ratio of the length of the DW to the transverse distance where the DW is annihilated. It is interesting to note that the range of the variation of the exponent $\tau_s$, induced by these geometrical effects, coincides with the experimental results in
  \cite{koreans_JApplPhys2008}, despite the absence of dipolar  interactions. 
A summary of all scaling exponents is given in Table\ \ref{tab-exponents}.
\end{itemize}

\begin{table}
\label{tab-exponents}
\caption{Scaling exponents $\tau_s$ and $\tau_T$ of the domain-wall slip avalanche size and duration in Figs.\ \ref{fig-dspt-A}-\ref{fig-dspt-C}, determined in the beginning (HLB) part and integrated  (HLI) along the  hysteresis loop for different sample shapes indicated by A, B and C in Fig.\ \ref{fig-schema}.  The numerical error bars are shown in scopes. The corresponding anisotropy exponent $\zeta$, the fractal dimension $D_{||}$ and the dynamics exponent $z$ are computed using the scaling relations, see text. }
\begin{tabular}{|l|cc|cc|cc|}
\hline
exp\ \ sample& A:HLB & A:HLI &    B:HLB& B:HLI&      C:HLB& C:HLI\\
\hline
$\tau_s$ & 1.5(1)&1.44(7)  & 1.5(5) &1.33(4)   &1.5(6) & 1.39(7) \\
$\tau_T$ & 1.9(8)&1.69(4)  & 1.86(9) & 1.54(6)  &1.9(8) & 1.63(9) \\
\hline
$\zeta$ &0.92 & 0.55 & 0.61 & 0.62  &0.75 & 0.60 \\
$D_{||}$ &1.92 &1.55  & 1.61 & 1.62  & 1.75& 1.60\\
$z$ &1.84 & 1.11 & 1.22 &  1.23 &1.51 & 1.21 \\
\hline
\end{tabular}
\end{table}

The impact of the DW length for the scaling exponents of the
avalanches can be related with effective long-range correlations of disorder along the DW in a low-dimensional geometry. Potentially, such a DW ``elasticity'' can also be seen as the origin of Tsallis distributions in HLC avalanches. 
In the literature, different universality classes of disorder-induced critical behaviour have been revealed 
\cite{cieplak_optpath1PRL994,BT_PhysA1999}, corresponding to weak and strong disorder. The models with the long-range correlations of randomness  (quenched and annealed) have been studied by the renormalization group (RG) methods in \cite{BT_disorderinducedcrit1998,SOC-RG-Antonov2016}.
In this context, the anisotropy induced by the built-in DW manifests itself in the dynamics and the avalanche geometry, leading to the following spatial scaling of the avalanche size 
$s(x)=x_{||}^{D_{||}}\Phi(x_\perp/x_{||}^\zeta)$. Here, $\zeta$ identifies the anisotropy exponent and $D_{||}$ is the fractal
dimension of the avalanche in the direction parallel to the
wall. Using the RG arguments \cite{BT_disorderinducedcrit1998} for the diffusion in the presence of anisotropic correlations of disorder, one can derive these exponents from the avalanche statistics, in particular, $D_{||}\equiv
1+\zeta =\frac{\tau_T-1}{\tau_s-1}$ and, to leading order of RG,
$z=2\zeta$ \cite{BT_disorderinducedcrit1998}. The numerical values of the derived exponents are also listed in Table\ \ref{tab-exponents}, subject to the statistical error bars of the avalanche distributions. The largest difference in the anisotropy
exponent at HLB and HLC occurs in sample A with the long
DW. The difference is reduced in the symmetrical shape, sample C, while it practically vanishes in sample B with the small DW. A similar observation extends to the dynamical exponents $z$, cf.\ Table\ \ref{tab-exponents}. The two values of $z$ for sample A once again confirm the conclusions of two different types of stochastic processes occurring in HLB and HLC, suggested  by the multifractal analysis in Section\ \ref{sec:dynamics}. 

In summary, our study of the dynamics of injected DW in low-dimensional ferromagnets driven by the external field reveals its multifractal features and the dependence on the sample shapes, which can be relevant for practical applications, e.g., to design on the DW-based memory devices. Furthermore, these observations shed light to the dynamic criticality of the RFIM at its lower critical dimension. The observed dynamic critical behaviours reveal sensitivity to the boundary conditions, sample shapes and the DW length. These conclusions are derived from numerical simulations for moderately high random field pinning above the critical disorder. The case of the weak disorder is left for future work as well as further theoretical analysis by RG approaches,  which can precisely identify the potential universality classes.

\acknowledgments
The author acknowledges the financial support from the Slovenian
Research Agency (research code fund- ing number P1-0044).

%\bibliographystyle{unsrt}
%\bibliography{bhn_biblio.bib}

\end{document}